\documentclass[11pt,twoside]{article}
\usepackage{asp2010}

\begin{document}

\title{Observations of dusty torii and compact disks around evolved stars: the high spatial resolution infrared view}
\author{Olivier~Chesneau
\affil{H. Fizeau, Univ. Nice Sophia Antipolis, CNRS UMR 6525, Observatoire de  la C\^{o}te d'Azur, Av. Copernic, F-06130 Grasse, France}
}

\begin{abstract}
The recent high angular resolution observations have shown that the transition between a globally
symmetrical giant and a source surrounded by a spatially complex environment occurs relatively early, as
soon as the external layers of the stars are not tightly bound to the core of the star anymore. In this review, the emphasis will be put on the delineating the differences between the torus and disk classification through the presentation of many examples of near-IR and mid-IR high angular resolution observations. These examples cover the disks discovered in the core of some bipolar nebulae, post-AGB disks, the dusty environment around born-again stars and recent novae, and also the disks encountered around more massive evolved sources. We discuss the broad range of circumstances and time scales for which bipolar nebulae with disks are observed.\\
\noindent{\bf Keywords.}\hspace{10pt}Planetary Nebulae -- Infrared interferometry -- Stars: evolution
\end{abstract}

\section{Preamble: Direct detection of binarity with high spatial resolution imaging}
A major breakthrough on the study on asymmetrical Planetary Nebulae (PNs) has been the recognition by a large part of our community of the growing importance of binary systems as main shaping agent of the bipolar and asymmetrical PNs. Companions encompassing a large range of mass, from the stellar objects to jovian-mass planets are suspected to deeply influence the ejecta when the star reaches the AGBs or even as early as the RGB \citep{2009PASP..121..316D}. This interaction can potentially influence dramatically the fate of the star, leading to poorly known evolutionary paths, influencing deeply the time scales of the different evolutionary stages, eventually bearing only little similarities compared to the time scales involved for the evolution of a single, naked star \citep{2007BaltA..16..104F,2007BaltA..16..112V}. Nevertheless, it is still far from excluded that a single star may also provide the conditions for the shaping of an asymmetrical nebula, if it expels at one time or another some mass with a significant pole-to-equator density gradient \citep{2002ARA&A..40..439B}. 

Many detections of companion of PNs central stars were reported during this conference \citep[De Marco, Hrivnak, Hajduk, Miszalski...these proceedings]{2010NewA...15..483B, 2009A&A...496..813M} some based on difficult long-term spectral monitoring, and many on the recent extensive photometric campaigns from automated telescopes. The high spatial resolution techniques may at first sight represent a large potential for detecting some well-separated  (2-100\,milli-arcsecond, hereafter mas) companions, but they suffer from intrinsic constraints that do not allow them to currently play a significant role in that domain. Adaptive optics on 8m-class telescopes are mostly limited in terms of spatial resolution, whilst optical interferometry has strong limitations in terms of contrast and imaging capabilities. The brighter the circumstellar environment and the central source, the harder will be the detectability of a small point-like structure in the close vicinity. The best case is the detection of low $Teff$ companions ($\Delta \sim$5mag, d$\sim$10-100mas) around hot, and preferentially naked sources for which it is easy to separate the SED of the stellar components. For a review of the methods used for detecting binarity, the reader is referred to \citet{2008AIPC.1057....1J}.

\section{Defining the different kind of equatorial over-densities}
Whatever the origin of the shaping process of an asymmetrical planetary nebula, an equatorial over-density of material is often involved in the models, and a growing number of observations unveil their presence at many stages of the star evolution. A wealth of new high spatial resolution techniques are now routinely available for the observer, namely the adaptive optics techniques, the speckle and lucky imaging techniques (often called the 'burst modes') in the optical and the infrared, and also the interferometry in the optical, infrared, millimetric and radio wavelengths. All these techniques have their own spatial resolution, sensitivity, contrast and astrometric constraints to the point that it is often quite difficult to compare and put into context the outcome of these observations.
Hence, the vocabulary can often be somehow misleading. The detection of disks around evolved sources, as in  post-AGB stars, may be closely associated with a high probability to see a binary system. However, this claim is highly dependent on the kind of structure encompassed in the 'disk' denotation, and one shall in the following sections precise further the equatorial overdensities, dividing them into the 'torus' (or 'outflowing wind'), and the 'stratified disks' families. The detection of compact, hot, dust-less accretion disks whose SEDs peak in the UV/B and that are hardly detectable by the above mentioned infrared techniques is out of the scope of this review. 

\subsection{Dusty torii: wind related equatorial overdensities}
An equatorial overdensity is a region of higher density and lower expansion speed compared to the polar circumstellar regions. The kinematics of such a structure is dominantly radial, and the total angular momentum carried by the structure is limited. Equatorial overdensities are short-term structures (these are deflected winds), e.g. if the mass-loss ceases, then the fate of the material is to rapidly expand and vanish. The increase of density toward the equatorial plane can be ascribed easily by a dependence on the co-latitude of the star in spherical coordinates, and there is no mass nor energy storage in the structure.

To date, the best examples of such torii originate from millimetric interferometry that associates a good spatial resolution ($\sim$1") and an excellent spectral resolution (i.e. 1\,km.s$^{-1}$). One can cite as good example of expanding torii, \citet{2008ApJ...673..934D} or \citet{2007A&A...473..207P}. In the near- or mid-IR, it is more difficult to access the density distribution of the circumstellar material due to opacity effects. Nevertheless, if the scale height of the structure is comparable to its radial extension, without clear sign of marked density increase toward the equatorial plane, the torus hypothesis can be favored \citep{2009A&A...503..837V,2006A&A...448..203L}. Long slit spectroscopy can provide further evidence for this classification when a significant expansion velocity (i.e. $\sim$15-40 km.s$^{-1}$) of the structure can be measured \citep{2010MNRAS.401..405J,2010arXiv1006.5873J}.

\subsection{Stratified disks}
A disk exhibits a clear vertical stratification whose scale height is governed by the gas pressure only. Its aperture angle is very small (i.e. less than $\sim$10 degrees) and its kinematics is dominated by (quasi) Keplerian velocities, with a small expansion component ($\leq$10\,km.s$^{-1}$).  Thus its lifetime is much longer than structures blown within a wind, competing or even exceeding the reference time scale of a PN lifetime of a few tens of thousand years \citep{2003ARA&A..41..391V}. 

The density structure of a Keplerian disk is best ascribed in cylindrical coordinates with a radial law in the equatorial plane, and a vertical stratification perpendicular to this plane, for the hydrostatic equilibrium.

Such a model is used in a wealth of astronomical contexts, and in particular for the formation of Young Stellar Objects (YSOs). Some examples of application of a similar model can be found in this list: \citet{2009A&A...500.1065D,2008A&A...491..809F,2003ApJ...588..373W,2003A&A...398..607D}.

The density law used in this model is
\begin{equation}
	\rho\left(r,z\right)=\rho_{0}\left(\frac{R_{*}}{r}\right)^\alpha \exp{\left[-\frac{1}{2}\left(\frac{z}{h\left(r\right)}\right)^2\right]}
\end{equation}
\begin{equation}
	h\left(r\right)=h_{0}\left(\frac{r}{R_{*}}\right)^\beta
\end{equation}
\noindent
where {\it r} is the radial distance in the disk's mid-plane, $R_{*}$ is the stellar radius, $\beta$ defines the flaring of the disk, $\alpha$ defines the density law in the mid-plane and $h_{0}$ is the scale height at a reference distance (often 100 AU)  from the star.
Until recently, such Keplerian disks could only be studied by millimetric interferometry at the highest possible spatial resolution and for the closest targets. One of the best example of such disks is the Red Rectangle \citep{2009ApJ...693.1946W,2005A&A...441.1031B}. Again, the great advantage of such a technique is to provide both the spatial distribution and the kinematics of the source \citep{2007A&A...468L..45B}. 

In absence of any information on the kinematics in the infrared due to the limited spectral resolution of optical interferometric observations, one has to rely on some indirect evidence to claim for the discovery of a stratified dusty disk, namely one has to prove at least that the disk opening angle is small, or better, that the density law follows radially and vertically the model described above. This is a possible task when the disk is seen close to edge-on. In this case, the radial and vertical directions are well separated on the sky.
The best example is the discovery of the edge-on disk in the core of Menzel 3, very well fitted by a stratified disk model \citep{2007A&A...473L..29C}. However, our group has conducted some tests trying to invert the disks parameters from artificial interferometric datasets (Niccolini et al., A\&A, accepted). When the disk is seen at low inclination, many degeneracies appear between the parameters of the vertical density law (such as the flaring parameter $\beta$), and the parameters of the radial law. Moreover, the size of the dust grains and and their composition affect also critically the fits.

To improve the constraints on the disk temperature law, a good approach is to perform near-IR and mid-IR observations using the AMBER and MIDI instruments of the VLTI respectively. The post-AGB binary IRAS 08544-4431 was studied this way \citep{2007A&A...474L..45D} and the same strategy is used for YSOs \citep{2008ApJ...676..490K}. It is more difficult to model the near-IR interferometric data due to the potential spatial complexity of the dusty disk's inner rim, and an intense theoretical and observational effort is currently performed, mainly in the YSOs community to better understand the so-called 'puffed-up inner rims' \citep{2010arXiv1006.3485D,2009A&A...506.1199K,2007ApJ...661..374T}. A better determination of the spatial properties and fine chemical content of the dust forming region is a challenge for the future. Other interesting targets are the double-chemistry sources such as BM\,Gem, in which a companion has recently been discovered \citep{2008ApJ...682..499I, 2008A&A...490..173O, 2006A&A...445.1015O}. These sources seem to harbor systemically an equatorially enhanced circumstellar environment, but there is no firm confirmation yet that the structure is best described by a torus or a disk.

\section{Disk evolution}
\subsection{Stratified disk and binarity}

Stratified disks detected in evolved systems are potentially highly correlated with binaries as demonstrated by Van Winckel and collaborators on the environment of binary post-AGBs \citep[Gielen et al. these proceedings]{2008A&A...490..725G,2007BaltA..16..112V,2006MmSAI..77..943V,2006A&A...448..641D}. Grain growth, settling, radial mixing and crystallization are efficient in such an environment and the circumbinary disc of these sources seems to be governed by the same physical processes that govern the proto-planetary discs around young stellar objects. It seems that another distinctive character of these long-lived stratified disks as seen in the infrared would be their content in highly processed grains \citep{2010A&A...510A..30M,2008A&A...490..725G, 2006A&A...451..951I}.
The key point for the stabilization of the disk is to provide enough angular momentum \citep{2008NewA...13..157A, 1997ApJS..112..487S}. A star may (hardly) supply this angular momentum via magnetic fields \citep{2004ApJ...614..737F, 2002ApJ...576..413U}. But even in this case the formation and stabilization of a Keplerian disk remains a challenge. Of course, this argument does not apply to the accretion disks encountered around YSOs, for which the difficulty in the contrary is to understand how the excess of angular momentum is dissipated.
The angular momentum provided by a low-mass stellar or even sub-stellar companion may potentially have a dramatic influence on a growing RGB or AGB star \citep{2010arXiv1002.2216N, 2010NewA...15..483B, 2007MNRAS.376..599N, 2006MNRAS.370.2004N,2000ApJ...538..241S}. The presence of a stratified disk, and the associated Lindblad resonances, seem to be a key ingredient in the orbital evolution of the binary system. Without this ingredient, it is difficult to reproduce the observed morphology of the eccentricity - period diagram (Dermine T. et al., in press)

{\it My personal opinion is that the discovery of a stratified disk with proven Keplerian kinematics is directly connected to the influence of a companion, albeit the few exceptions presented above, namely the Young Stellar Objects or the critical velocity rotating massive sources. This hypothesis must be confirmed by further observations. }

\subsection{Time scale of formation}
When dealing with the theoretical building-up of a disk, time matters, and any constraint on the time-scale on which the observed structures were formed is of importance. An interesting study was presented by \citet{2007ApJ...663..342H} on the close time-scale connection between jets and torii, based on many kinematical studies. $\pi^1$ Gruis is a good example of a recently formed structure \citep{2008A&A...482..561S, 2006ApJ...645..605C}. OH231.8+4.2 is another example of a similar process caught in the act in IRAS16342-3814 \citep{2010arXiv1004.2659E, 2006ApJ...646L.123M,2007A&A...468..189D}. 

Some recent examples show how fast may be the building up of an equatorial overdensity. 
The born-again stars, V605\,Aql \citep{2008A&A...479..817H} and the Sakurai's Object are surrounded by an equatorial overdensity with a large scale height that may be described as a torus \citep{2009A&A...493L..17C}. In the case of the Sakurai's object, the torus was detected in 2007, about 10 years after the dust began to form. A dense equatorial structure was formed even faster, in less than 2yrs, around the slow dust-forming nova V1280 Sco \citep[Chesneau et al. in preparation]{2008A&A...487..223C}. There is no doubt in this case that the slow ($\sim$500km.s$^{-1}$) ejecta from the outbursting nova were deeply affected by the common envelope phase that lasted more than tens of orbital period of the companion, leading to the fast formation of a bipolar nebula (see also Evans, these proceedings).

These examples show that as soon as the primary in a binary system gets larger and increases its mass-loss rate when evolving, the influence of a companion can rapidly focus the ejected material onto the equatorial plane of the system, leading to an equatorial overdensity, as already investigated theoretically \citep{1998ApJ...497..303M,1993MNRAS.265..946T}. This physical process depends on many parameters (mass ratio, orbital parameters, mass-loss rate of the primary...), but the efficiency is such that a large number of targets might be affected at one stage or another \citep{2010NewA...15..483B}.

\subsection{Time scale of dissipation and fate}
The observations of the inner circumstellar structures around evolved sources have been to date too scarce to put them into an evolutionary sequence. As written above, the torii are supposed to expand and dissipate much faster than the Keplerian disks. One may even consider the extreme case in which a binary system surrounded by a stable circumbinary disk continuously replenished by the interaction of the stars may remain virtually unchanged for time scale as long as many 10$^5$-10$^6$years, as proposed for the Red Rectangle. A proposed unified picture of the such an interacting binary evolution involving the presence of a torus is presented in \citet{2007BaltA..16..104F}.

What is the fate of a stratified disk? An 'old' dissipating disk can see its density and kinematical structure deeply affected by the fast evolving wind of the central star. \citet{2010A&A...514A..54G} performed an in-depth kinematical study of the dissipating disk in the core of the M2-29 nebula \citep{2008A&A...490L...7H}. The torus found around Hen2-113 might also be dissipating under the influence of the fast radiative wind emitted by the Wolf-Rayet star in the core of the nebula \citep{2006A&A...448..203L}.

\section{Stratified disks around massive stars}
There is now a large bunch of evidence that the disks encountered around Be stars and at least some B[e] supergiants are rotating close to Keplerian velocities \citep{2007A&A...464...59M}. The Be stars are proven statistically to represent the tail of the fastest rotators among B stars, and their disks of plasma may be explained without invoking the influence of a companion, by a combination of an extreme centrifugal force at the equator and some pulsation properties of the star which leads to erratic ejection of material with high angular momentum \citep{2009ApJ...701..396C, 2009A&A...506...95H}. By contrast, there is still no consensus for the more massive supergiant counterparts, the B[e] stars that are surrounded by dense disks of plasma {\it and} dust. The dust survives much closer to this hot star than expected so far  \citep{2010A&A...512A..73M, 2009A&A...507..317M, 2007A&A...464...81D}.The B[e] supergiant's rotation rate is strongly decreased by the increase of their radius while leaving the main sequence, and the rotation alone is probably far from being sufficient to explain the formation of a disk without invoking the influence of a close companion \citep{2007ApJ...667..497M}. Note also, that dusty circumbinary disk are commonly encountered around interacting binary systems \citep{2009A&A...499..827N}.

The example of the A[e] supergiant HD\,62623 is informative in this context \citep[Millour, Meilland et al., submitted]{2010A&A...512A..73M}. HD\,62623 is an A supergiant showing the characteristics of the 'B[e]' spectral type, namely a spectrum dominated by strong emission lines and a large infrared excess. Spectrally and spatially resolved observations of AMBER/VLTI in the Br$\gamma$ line have shown that the supergiant lies in a cavity, and is surrounded by a dense disk of plasma. The Br$\gamma$ line in the location of the central star is {\it in absorption} showing that the star is not different from a normal member of its class such as Deneb (A3Ia), albeit with a significantly large v$sini$ of about 50km.s$^{-1}$. By contrast, the Balmer and Bracket lines are wider ($vsini$ of about 120km.s$^{-1}$), and the AMBER observations demonstrated that they originate from a disk of plasma, most probably in Keplerian rotation (Millour, Meilland et al. submitted). In absence of any proof of binarity, it is often difficult to understand how such a dense equatorial disk could have been generated. However, HD62623 is a known binary with a stellar companion that orbits close to the supergiant with a period of about 136 days \citep{1995A&A...293..363P}. The mass ratio inferred is very large, and the companion is probably a solar mass star. \citet{1995A&A...293..363P} proposed that an efficient angular momentum transfer occurs near the L2 Lagrangian point of the system, propelling the mass lost from the supergiant by its radiative wind and probably also by strong tides into a stable dense circumbinary disk \citep[in the context of an AGB star]{2009A&A...507..891D}. A similar idea was proposed by \citet{2009A&A...507..891D} in relation with radiation pressure acting on the wind of AGB stars and modifying the Roche lobe geometry, therefore probably easing the formation of a circumbinary disk.
 
The comparison between the disks encountered around low and intermediate mass stars and those observed around the B[e] supergiants, a rare spectral type among the zoo of massive stars might not appear relevant at first sight. Yet, recent Spitzer observations of 9 LMC B[e] stars showed an interesting homogeneity of their spectra, and a great similarity with the spectra of post-AGBs harboring dense dusty disks \citep{2010AJ....139.1993K}. This tightens further the connection between B[e] stars and binarity. 

\acknowledgements This article benefited from fruitful discussions with many persons, and in particular Alain Jorissen.

\bibliography{Biblio_disks}

\end{document}